\documentclass[twocolumn,showpacs,preprintnumbers,amsmath,amssymb,aps,prl]{revtex4}

\usepackage{graphicx}
\begin{document}

\title{Jamming in Granular Polymers}   
\author{L. M. Lopatina,    
 C. J. Olson Reichhardt, and C. Reichhardt} 
\affiliation{
Theoretical Division,
Los Alamos National Laboratory, Los Alamos, New Mexico 87545} 

\date{\today}
\begin{abstract}
We examine the jamming transition in 
a two-dimensional granular polymer system using compressional
simulations. The jamming density 
$\phi_{c}$ decreases with increasing length of the granular chain
due to the formation of loop structures,
in excellent agreement
with recent experiments. The jamming density 
can be further reduced in mixtures
of granular chains and granular rings, also as observed in experiment. 
We show that the nature of the jamming
in granular polymer systems 
has pronounced differences from the 
jamming behavior observed for polydisperse 
two-dimensional disk systems at Point J. This result provides
further evidence
that there is
more than one type of jamming transition.   
\end{abstract}
\pacs{45.70.-n,81.05.Rm,83.80.Fg}
\maketitle

\vskip2pc
Jamming, or the development of a resistance to shear,
is a phenomenon that occurs when a disordered
assembly of particles subjected to increasing density, load 
or other perturbations exhibits a transition from a 
liquid-like state that can flow to a rigid state that acts
like a solid under compression.
Tremendous recent growth in this field has been driven 
by the prospect that jamming may be associated with universal properties 
across a wide class of systems 
including granular media, foams, emulsions,
colloids, and glass forming materials \cite{Lui}. 
One of the most accessible routes for exploring the jamming 
transition is gradually increasing the density of a sample in the
absence of shear or temperature.  Here, jamming occurs at a density
termed `Point J.' \cite{Silbert}.
Jamming transitions have been studied both for noncohesive granular
media and for cohesive and/or nonspherical granular materials
\cite{Gilabert,Lois,Mailman,Donev}.
There is considerable evidence that for frictionless
disordered disk assemblies, critical behavior occurs near
Point J, with both the pressure and the particle coordination number
$Z$ exhibiting power law behavior as a function of packing
density $\phi$ 
\cite{Silbert,Hern,Senke,Sperl}.  
Similar behavior appears 
when the shear, external forcing, or temperature are finite, 
providing further evidence that 
the jamming transition may indeed exhibit universal properties
\cite{Teitel,B,Hastings,Durot,Head,Durian,Zhang}.  
If such universal behavior holds
for other systems that undergo jamming, 
it would have profound implications for the
understanding and control of disordered and glassy systems.  

The most widely studied two-dimensional (2D) jamming system contains
bidisperse
frictionless disks.  When two sizes of disks 
with a radius ratio of $1:1.4$ are mixed in a 50:50 ratio, a jamming
transition occurs at a density of $\phi = 0.84$ 
\cite{Silbert,Sperl,Teitel,B,Hastings,Head,Zhang}.   
To explore whether the jamming transition is universal in nature, it would
be ideal to have a system in which the jamming density $\phi_c$ could
be tuned easily.
Here we propose that one model system which meets this criterion is
an assembly of 2D granular polymers.
This model 
is motivated by experiments on granular polymers or chains 
of the type used for lamp pulls,
where various aspects of knot formation, diffusion processes, and
pattern formation have been explored   
\cite{Ecke,Safford,Mullin}. 
We model the chains as coupled harmonically repulsive disks 
similar to those studied in the polydisperse
disk system, with a constraint on the minimum angle that can be spanned
by a string of three disks.
Other workers have considered freely-jointed chains
\cite{karayiannis} or chains of sticky spheres \cite{hoy}.
Although our model is 2D and neglects friction, we show
that it captures the same features found in recent 
three-dimensional (3D) granular polymer compaction
experiments \cite{HJ,Olson}. 
To study the jamming transition, 
we construct pressure versus $\phi$ curves
by compressing the chains between two walls, and compare our results
to compression experiments on 2D polydisperse disks
\cite{Sperl,Ecke}.
We show that the jamming density $\phi_c$ decreases  
with increasing chain length and 
saturates at long chain lengths, in agreement with
the experiments of Ref.~\cite{HJ}. 
The decrease of $\phi_c$ occurs 
due to the formation of rigid loops along the chains
which stabilize voids inside the packing. 
Unlike the bidisperse 
disk system where the pressure scales nearly linearly with $\phi$,
in the chain system the pressure increases 
with a power law form with an exponent significantly larger than 1
as the jamming transition is approached.    
Jamming of our frictionless granular chains shares several features
with jamming of frictional disks and could be distinct from the jamming
transition for frictionless disks.
         
\begin{figure}
\includegraphics[width=3.5in]{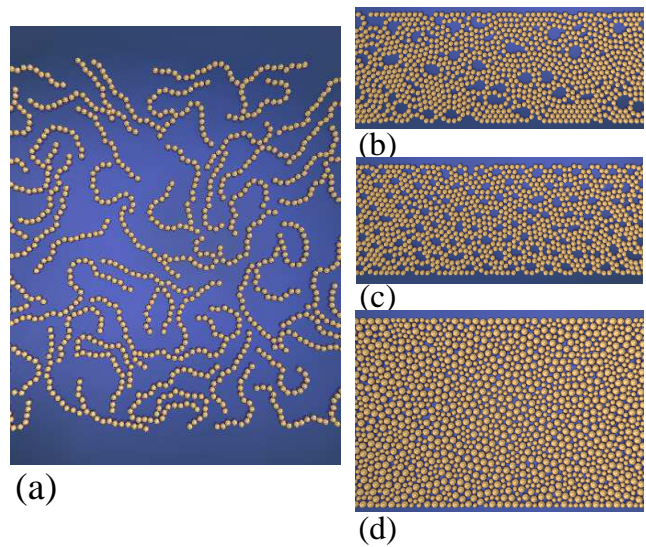}
\caption{
(Color online) Granular configurations in a portion of the sample.  The $x$
direction runs vertically and the fixed wall is at the
top of each panel.
(a) An unjammed system of 
$N=67$ granular polymers with length $M_{b} = 16$.
(b) The jammed configuration for the same system contains
voids which appear when the chains form ring structures.   
(c) The jammed configuration for $N=67$ loops of 
length $M_{b} = 16$. 
The jamming density is lower than for 
systems of
chains or individual disks.
(d) The jammed configuration 
at $\phi=0.84$ for a sample of $N=1500$ bidisperse frictionless disks
contains no significant voids.
}
\label{fig:1}
\end{figure}

We simulate a 2D system confined by two walls at $x=0$ and
$x=L$ and with periodic boundaries in the $y$ direction.  The wall at $x=L$
is held fixed while the position of the other wall is allowed to vary in order
to change the density.
The system contains $N$ chains or loops, 
each of which is composed of $M_b$ individual
disks that are strung together by harmonic springs and that experience a
constraining force which limits the bending radius of the chain.
In loops, the two ends of a chain are connected together.
The disk-disk interaction is modeled as 
a stiff harmonic repulsion, and the motion of all disks is taken to be 
overdamped in order to represent the frictional force between 
the disks and the underlying floor.
A given disk $i$ moves according to the following equation of motion:      
\begin{equation}
\eta\frac{d {\bf R}_i}{dt} = {\bf F}_{dd}^i + {\bf F}_{c}^i + {\bf F}_{cc}^i 
+{\bf F}_w^i.
\end{equation}
Here we take the damping constant $\eta=1$.
The disk-disk interaction potential is
${\bf F}_{dd}^i=\sum_{j\ne i}^{NM_b}k_g(r_{\rm eff}-r_{ij})\Theta(r_{\rm eff}-r_{ij})
{\bf \hat r}_{ij}$
where the spring constant $k_g=300$, $r_{ij}=|{\bf R}_i-{\bf R}_j|$,
${\bf \hat r}_{ij}=({\bf R}_i-{\bf R}_j)/R_{ij}$, $\Theta$ is the
Heaviside step function, and $r_{\rm eff}=r_i+r_j$, where $r_{i(j)}$ is the
radius of disk $i(j)$.  For the chains and loops
we set $r_i=1$; for a bidisperse disk
system we set $r_i=1$ for half of the disks and $r_i=1.4$ for the other half
of the disks.
The chain interaction potential is
${\bf F}_{c}^i=\sum_{k}k_g(r_{\rm eff}-r_{ik}){\bf {\hat r}}_{ik}$, 
and it acts only between
a disk and its immediate neighbors along the loop or 
chain.
The bending constraint potential
${\bf F}_{cc}^i=\sum_{l}k_{g}(r_{\rm stiff}-r_{il})\Theta(r_{\rm stiff}-r_{il})
{\bf {\hat r}}_{il}$ 
acts
between disks separated by one chain element, with 
$r_{\rm stiff}=2r_{\rm eff}\sin(\theta_s/2)$ 
and $\theta_s=0.82\pi$ unless
otherwise noted.
Smaller values of $\theta_s$ produce more bendable chains.
The disk-wall interaction force ${\bf F}_w^i$ is computed by placing a
virtual disk at a position reflected across the wall from the actual disk,
and finding the resulting disk-disk force.
To initialize the system, we place the chains, rings, or 
individual disks in random non-overlapping positions 
to form a low density unjammed phase,
such as in Fig.~\ref{fig:1}(a). 
The $x=L$ wall is held fixed while 
the other wall is gradually moved from $x=0$ to larger $x$
in small increments of $\delta x$. 
The waiting time between increments is taken long enough
so that the system has sufficient time 
to relax to a state where 
the velocities of all particles are indistinguishable from zero.

We identify the jamming transition by measuring the total force exerted on
the fixed wall by the packing,
$P=\sum_i^{NM_b}{\bf F}_w^i \cdot {\bf \hat x}$, 
and the average contact number $Z=(NM_b)^{-1}\sum_i^{NM_b}z_i$ 
as a function of the total 
density of the system defined by the spacing between the two walls.
To determine the contact number $z_i$ of an individual grain in a chain,
we first count the immediate neighbors of the grain along the chain,
and then add any other grains that are in contact with the
individual grain.
The force $P$ is proportional to the $p_{xx}$ component of the pressure tensor.
At the jamming transition, 
the pressure in the system becomes finite \cite{Silbert,Hern,Sperl}, while
below jamming $P=0$.   
Previous simulations on 2D disordered disk 
packings have revealed the onset of a finite pressure  
near $\phi_{c} = 0.84$ which  
grows as 
$P \propto (\phi - \phi_{c})^{\psi}$ with $\psi = \alpha_{f} -1$, where
$\alpha_{f}$ is the exponent of the interparticle interaction potential 
\cite{Silbert,Hern}. 
Recent work which includes careful corrections to scaling indicates that
the exponent $\psi=1.1$ \cite{2011}.
Theoretical work on the jamming of 2D disks
also predicts a power law scaling of the
pressure versus density \cite{Senke}, and indicates that 
the contact number $Z$ 
should scale as $Z \propto (\phi - \phi_{c})^{\beta}$, with $\beta = 0.5$.  
Experiments using a combination of shear and compression 
on the same disk system
found that 
after performing cycling to reduce the effect of friction,
the pressure and Z both scale with the density as power laws 
with $\psi = 1.1$ and $\beta = 0.495$ \cite{Sperl}.   

\begin{figure}
\includegraphics[width=3.5in]{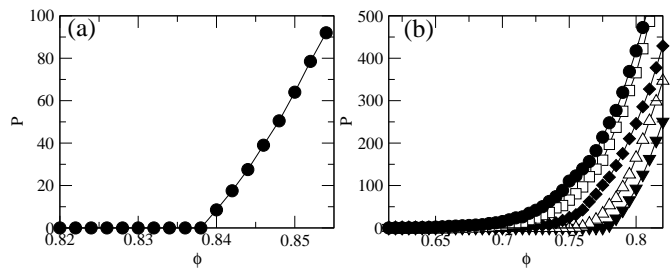}
\caption{
(a) The pressure $P$ vs $\phi$ 
for a bidisperse disk system. 
Above the jamming transition at $\phi\approx 0.84$, $P$ increases
nearly linearly with $\phi$.
(b) $P$ vs $\phi$ for the granular polymer system 
with chains of length
$M_b=6$ ($\blacktriangledown)$,
8 ($\bigtriangleup$),
10 ($\blacklozenge$),
16 ($\square$),
and 24 ($\bullet$).
As $M_{b}$ increases, the onset of a finite value of $P$ indicating jamming
drops to lower values of $\phi$ and the near linear scaling of $P$ with $\phi$
is lost.
}
\label{fig:2}
\end{figure}

\begin{figure}
\includegraphics[width=3.5in]{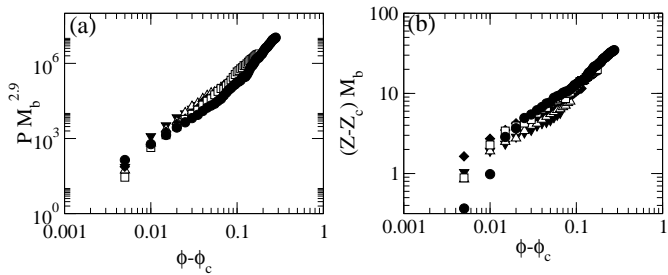}
\caption{
(a) Scaling of pressure vs density, 
$P M_b^{2.9}$ vs $\phi-\phi_c$ close to the jamming transition for
chains of length
$M_b=6$ ($\blacktriangledown)$,
8 ($\bigtriangleup$),
10 ($\blacklozenge$),
16 ($\square$),
and 24 ($\bullet$).
(b)
Scaling of $(Z-Z_c) M_b$ vs $\phi-\phi_c$ for chains of length
$M_b=6$, 8, 10, 16, and 24, with the same symbols as in panel (a).
}
\label{fig:scale}
\end{figure}

\begin{figure}
\includegraphics[width=3.5in]{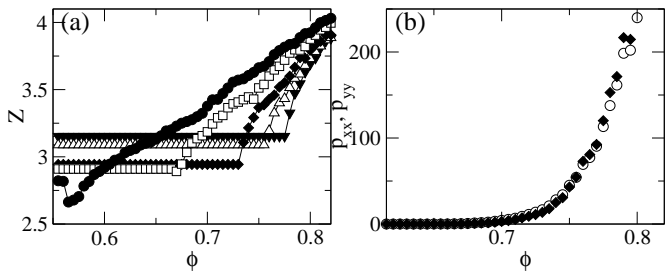}
\caption{
(a) Contact number $Z$ vs $\phi$ for chains of length
$M_b=6$ ($\blacktriangledown)$,
8 ($\bigtriangleup$),
10 ($\blacklozenge$),
16 ($\square$),
and 24 ($\bullet$).
(b) Bulk pressure tensors $p_{xx}$ 
($\bigcirc$)
and $p_{yy}$ 
($\blacklozenge$)
vs $\phi$ for $M_b=16$.
}
\label{fig:2B}
\end{figure}

\begin{figure}
\includegraphics[width=3.5in]{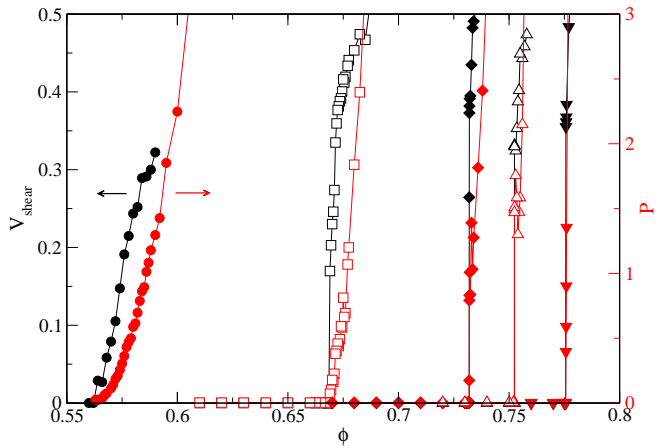}
\caption{
(Color online)
The shear velocity $V_{shear}$ of grains adjacent to the stationary wall vs
$\phi$ (dark black symbols) and the corresponding pressure $P$ in the packing vs
$\phi$ (light red symbols) for a sample in which a shear force is applied using
the mobile wall with chain lengths
$M_b=6$ ($\blacktriangledown)$,
8 ($\bigtriangleup$),
10 ($\blacklozenge$),
16 ($\square$),
and 24 ($\bullet$).
The onset of a finite pressure and a finite shear response occur at the
same value of $\phi$ for each sample.
}
\label{fig:shear}
\end{figure}

We first test our simulation geometry using the bidisperse individual
disk system.  A configuration of $N=1500$ disks appears in 
Fig.~\ref{fig:1}(d) just above the onset of a finite pressure $P$
at $\phi \approx 0.84$. 
In Fig.~\ref{fig:2}(a) we 
show that
for the disk system 
above jamming, $P$
increases nearly linearly with $\phi$, consistent with a scaling exponent
$\psi=1.1 \pm 0.1$.
This indicates that our compressional geometry captures the 
jamming behavior found in other studies of bidisperse disks.

We use the same compression protocol 
to study the jamming behavior of granular polymers, as illustrated 
in Fig.~\ref{fig:1}(a,b) 
for a system with $N=67$ chains that are each of length 
$M_b = 16$. 
In Fig.~\ref{fig:2}(b) we plot 
$P$ versus $\phi$ for granular polymer chains of lengths
$M_b=6$, 8, 10, 16, and 24.
For the chains, the onset of finite $P$ indicating the beginning of jammed
behavior 
occurs at a much lower density than for the 
bidisperse disk system shown in Fig.~\ref{fig:2}(a), and as $M_b$ increases,
the jamming transition shifts to even lower $\phi$ and the nearly linear
dependence of $P$ on $\phi$ is lost.
We illustrate scaling of $P$ near the jamming transition in
Fig.~\ref{fig:scale}(a) where we plot $P M_b^{2.9}$ vs $\phi-\phi_c$.
Here we find $\psi \approx 3.$
The jammed state develops isotropic rigidity, as indicated by the plot
of the bulk pressure tensor components $p_{xx}$ and $p_{yy}$ in 
Fig.~\ref{fig:2B}(b).
To test whether the packing also develops a finite response to shear at the
jamming transition, we fix the packing density and apply a shear to the system
by applying a force ${\bf F}_{shear}=0.04{\bf \hat y}$ to all particles that are
in contact with the mobile wall.  We measure the resulting 
steady state velocity 
$V_{shear}=d{\bf R}_i/dt \cdot {\bf \hat y}$ of all particles that are in
contact with the stationary wall on the other side of the sample, discarding
any brief initial transient responses.  In
Fig.~\ref{fig:shear} we plot $V_{shear}$ and $P$ versus $\phi$ for samples with
chains of length $M_b=6$, 8, 10, 16, and 24.  In each case, a finite shear 
response and a finite pressure $P$ appear simultaneously at the jamming
density.  Our use of a fixed $F_{shear}$ at all densities rather than a variable
shear rate affects the shape of the $V_{shear}$ versus $\phi$ curves; 
variations in the shear rate as well as related hysteretic effects under
shear will be the subject of a future publication.

\begin{figure}
\includegraphics[width=3.5in]{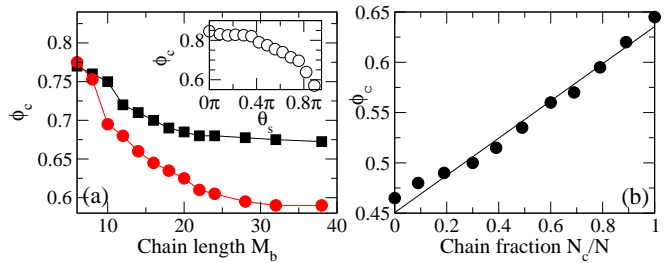}
\caption{ 
(Color online)
(a) The jamming threshold $\phi_{c}$ versus 
chain length $M_{b}$ for chains with a 
bending angle of 
$\theta_s=0.82\pi$ (circles) and $\theta_s=0.756\pi$ (squares).
$\phi_c$ decreases with increasing $M_{b}$ and 
saturates at large $M_b$. 
Inset: $\phi_{c}$ vs $\theta_s$ for a system with $M_{b} = 16$. 
(b) $\phi_{c}$ vs chain fraction $N_c/N$
for a system with $M_{b} = 16$ and a mixture of loops and chains. 
As the fraction of chains decreases, $\phi_{c}$ decreases. 
}
\label{fig:3}
\end{figure}

We define the jamming threshold $\phi_c$ as the density at which $P$ 
rises to a finite level.  The same threshold also
appears as the sudden onset of an increase in $Z$, as seen
in 
Fig.~\ref{fig:2B}(a).
The scaling of $(Z-Z_c) M_b$ vs $\phi - \phi_c$ appears in
Fig.~\ref{fig:scale}(b), where the exponent $\beta$ falls in
the range $\beta=0.6$ to 0.8.
In Fig.~\ref{fig:3} we plot $\phi_{c}$ versus $M_b$ for chains of two 
different stiffnesses: $\theta_s=0.82\pi$ and $\theta_s=0.756\pi$.
In both cases, $\phi_{c}$ decreases
monotonically with increasing $M_b$ and saturates
for large $M_b$.   
Recent experiments on the packing of granular polymers showed the same 
behavior: the final packing density decreased for increasing chain length
and saturated for very long chains 
\cite{HJ}.  
This was attributed to the formation of rigid semiloops 
which stabilized voids in the packing and decreased the jamming density.
Loops have also been observed in dense 3D packings of freely-jointed
chains \cite{preloop}.
Since the minimum area spanned by a semiloop
increases with $\theta_{s}$, the 
jamming density should be lower for larger
$\theta_{s}$ when larger voids are stabilized.
Figure \ref{fig:1}(b) illustrates the voids that appear in our chain packings due to
the formation of rigid semiloops.
For comparison, the bidisperse disk system shown in Fig.~\ref{fig:1}(d) contains no
large voids.
If the rattler disks in Fig.~\ref{fig:1}(d) were removed, the amount of void
space present would increase, but the chain system would still be able to
stabilize a larger amount of void space since the constraint of the
chain backbone permits the formation of larger arches around the voids
than the arches that can be stabilized in the bidisperse disk system.
The semiloops increase in size for increasing
$\theta_{s}$ and $\phi_c$ is reduced at higher $\theta_s$, as shown
in Fig.~\ref{fig:3}(a).
The plot of $\phi_c$ versus $\theta_s$ in the inset of Fig.~\ref{fig:3}(a) for 
fixed $M_b=16$ shows that
$\phi_{c}$ monotonically decreases with increasing $\theta_{s}$. 
For perfectly flexible chains with $\theta_s=0$, we find $\phi_c \approx 0.8$.
This is lower than the density of a triangular lattice due to the trapping
of voids within the packing by the physical constraints of the bonding between
chain elements.  If the system were annealed or shaken for a sufficiently
long time, these
voids could eventually be freed and the perfectly flexible chains would form
a perfect triangular lattice.

As a confirmation of the idea that the formation of rigid semiloops 
is the mechanism by which the jamming density is depressed, 
Ref.~\cite{HJ} includes experiments performed on mixtures of granular 
polymers and granular loops with equal length $M_b$.
In this case, $\phi_c$ decreased linearly as the fraction of loops
increased.
We find the same effect in our 2D system by
varying the number of loops $N_l$ and chains $N_c$ in a sample with
fixed $N=N_l+N_c$ and fixed $M_b$.
In Fig.~\ref{fig:3}(b) we plot $\phi_{c}$ versus $N_c/N$, where $N_c/N=1$ indicates
a sample containing only chains and $N_c/N=0$ is a sample containing only
loops.
As $N_c/N$ decreases, $\phi_c$ decreases.
The jammed configuration for a sample with $M_b=16$ containing only loops,
$N_c/N=0$, appears in Fig.~\ref{fig:1}(c).
The number of voids present is much larger than in the $N_c/N=1$ sample
shown in Fig.~\ref{fig:1}(b).
Interestingly, the voids began to form a disordered triangular packing.

Our results indicate that the experimentally observed dependence of the
jamming density on chain length or fraction of loops in Ref.~\cite{HJ}
is not caused by friction or other
possible spurious effects, 
but is instead a product of the 
geometrical configuration of the chains and loops. 
The surprisingly good agreement between our 2D simulations and the
3D experiments may be due to the fact that in each case,
the semiloops formed by the chains are 2D in nature.
Additionally, in the experiment the container used to hold the sample
induced ordering of the chains and loops near the walls and may have caused
the system to act more two-dimensional.
The fact that much of the physics observed for the 
3D system can be captured in 2D models 
means that 2D experiments, which are much easier to image than 3D
experiments, could provide many of the
same insights for understanding jamming 
in a system where the jamming density can be tuned easily.

\begin{figure}
\includegraphics[width=3.5in]{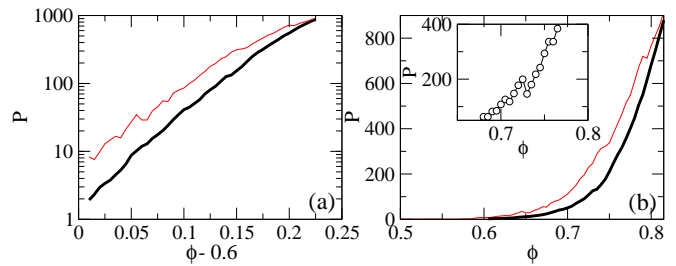}
\caption{
(Color online)
(a) $P$ vs $\phi$ for a chain
sample with $M_b = 20$ on a log-linear scale. 
Light red line: initial compression; heavy black line: steady state reached
after four compressions. 
(b) $P$ vs $\phi$ for a system with $M_{b}  = 38$ for the initial compression 
(light red line)
and for the fourth compression (heavy black line). 
The response is hysteretic but the curves remain nonlinear on all
compressions.
Inset: A portion of the $P$ vs $\phi$ curve during the second compression
cycle in the same sample showing a sudden pressure 
drop associated with the collapse of an unstable void. 
}
\label{fig:4}
\end{figure}

The chain system exhibits a pronounced hysteresis effect that can be seen by 
cycling the mobile wall in and out to 
repeatedly compress and uncompress the packing.
Figs.~\ref{fig:4}(a) and (b) show a comparison of the responses during the first
compression and during the fourth compression.  After four compressions
the system does not exhibit further hysteresis. 
In contrast, we find little or no hysteresis for the 
bidisperse disk system.
During the initial compression, 
the chain systems often exhibit sizable fluctuations in $P$
above the onset of jamming.
Sudden drops in the pressure,
such as that shown in the inset of Fig.~\ref{fig:4}(b), 
occur due to the collapse of semiloops that 
are larger than the minimum stable size. 
After all semiloops have reached a stable size, 
we find no further hysteresis. Even after cycling to a steady state, the
$P$ vs $\phi$ curves remain power law 
in nature with an exponent significantly larger than 1
and do not become linear or nearly linear. 
Simulations of compressed 2D frictional bidisperse disk systems 
show that 
void structures can form during the initial compression but 
collapse during subsequent cycles, allowing the sample to reach the same
density as a frictionless disk sample \cite{Lena}.
In the chain system, the void structures are associated with semiloops that
have formed in the chains and, unlike in the disk system, the voids can never
be fully collapsed by repeated cycling.

The fact that the granular polymers do not exhibit the same behavior at
the jamming transition as the bidisperse disk systems do at 
Point $J$ 
provides additional evidence
that jamming does not occur with universal features in all systems,
and that the criticality found in the
bidisperse disk systems may be associated with a special type of jamming. 
Additional studies on a variety of different types of systems 
would need to be performed to confirm whether
the jamming behavior is indeed different for each system
or whether there is a small number of 
different classes of jamming behaviors, with the granular polymer system and
the bidisperse disk system falling into separate classes.

In summary, we have introduced a numerical model of 2D granular polymers
that can be used to study the jamming transition.     
The onset of jamming occurs at a density that decreases with increasing
chain length and saturates for long chain lengths.
The decrease of the jamming density results from the formation of rigid
semiloops in the granular chains which permit stable voids to exist in the
packing,
in excellent agreement with recent 3D experiments on granular polymers. 
For fixed chain length, 
the jamming density decreases when the chains are made stiffer since the
rigid semiloops, and the voids stabilized by them, are larger.
The jamming density can also be further decreased
by increasing the fraction of granular loops present in the packing, which is
also in agreement with experimental observations.
The fact that our 2D simulations agree so well with the 
3D experiments of Ref.~\cite{HJ} indicates that the formation of semiloops in
the chains is essentially a 2D phenomenon.     
In comparison to bidisperse disk systems 
which show a nearly linear increase in the pressure as a function of density,
characteristic of a critical phenomenon, 
in the granular polymer systems the pressure increases 
as a power law with exponent significantly larger than 1,
suggesting that the jamming transition
in the granular chain system is different in nature from jamming in 
the bidisperse disks and may be related to the type of
jamming that occurs for frictional grains.

We thank R. Ecke and R. Behringer for useful comments.
This work was carried out under the auspices of the 
NNSA of the 
U.S. DoE
at 
LANL
under Contract No.
DE-AC52-06NA25396.

\end{document}